# Direct Measurement of the Radiative Pattern of Bright and Dark Excitons and Exciton Complexes in Encapsulated Tungsten Diselenide

Lorenz Maximilian Schneider[1,*], Shanece S. Esdaille[2], Daniel A. Rhodes[2], Katayun Barmak[3], James C. Hone[2], and Arash Rahimi-Iman[1,*]

[1]*Faculty of Physics and Materials Sciences Center, Philipps-Universität Marburg, Marburg, 35032, Germany*

[2]*Department of Mechanical Engineering, Columbia University, New York, NY 10027, USA*

[3]*Department of Applied Physics and Applied Mathematics, Columbia University, New York, NY 10027, USA*

**Abstract:** The optical properties of particularly the tungsten-based transition-metal dichalcogenides are strongly influenced by the presence of dark excitons. Recently, theoretical predictions as well as indirect experimental insights have shown that two different dark excitons exist within the light cone. While one is completely dark, the other one is only dipole forbidden out-of-plane, hence referred to as grey exciton. Here, we present angle-resolved spectroscopic data of a high-quality hexagonal-BN-encapsulated WSe$_2$ monolayer with which we directly obtain the radiation pattern of this grey exciton that deviates from that of the bright exciton and other exciton complexes obtained at cryogenic temperatures.

**Introduction**

Transition-metal dichalcogenides (TMDCs) have recently drawn lots of attention due to their extraordinary rich exciton physics[1], showing trions[2,3], biexcitons[4] and other higher-order exciton complexes[5,6]. Furthermore, caused by their huge binding energy, higher-order states of the Rydberg-like series can be even seen at room temperature[7,8]. In addition, even polaritons[9] and valley polaritons[10,11] can be observed in suitable microcavity structures at these temperature. With recent publications, it has been revealed that several dark excitons[12,13] exist that strongly affect the dynamics[14,15] and spectral features[13,16] of the system. While there have been several experimental studies for the dark exciton[17–20], little has been reported about their emission characteristics.

Recent detailed group-theory analysis of the possible excitons' branches[15] of WSe$_2$ has shown that four different exciton configuration exist within the light cone at the A-exciton peak (cf. **Figure 1**). Firstly, two bright excitons (Γ6) at K respectively K' are provided, where hole and electron exhibit the same spin. Secondly, owing to a small but nonnegligible intravalley interaction, two more possible states arise that represent a coherent superposition of intervalley excitons composed of the spin-forbidden transition across K and K' valley (irreducible representation Γ4 and Γ3). A schematic two-particle picture for these species is shown in **Figure 1a**, whereas their radiation pattern is indicated in **Figure 1b**. The Γ4 state transforms like a z-component of a vector of D$_{3h}$ group, rendering it a dipole-allowed transition for z-polarization (corresponds to the direction out of plane, as indicated by small arrows attached to the exciton representation in **Figure 1b**). In contrast, the Γ3 state is completely dark, i.e. it cannot couple to the electro-magnetic field.

This bears huge implications, as the Γ4 can only emit in the plane and not out of the plane. The energetic difference between bright and grey exciton will be here labelled Δ, whereas the one between grey and dark exciton will be represented by δ (c.f. **Figure 1a**).

This implication has already been investigated using a µ-PL setup with separate detection paths for optical x- and z- polarisation[17] for encapsulated samples. For a bare substrate-supported monolayer at room temperature, the radiation pattern of emission has also been addressed by Fourier-space spectroscopy[21]. However, to the best of our knowledge, no direct measurement of the radiation pattern has been performed for the dark exciton. Here, we present an angle-resolved photoluminescence study with the focus on the distinct radiation patterns of bright and dark states.

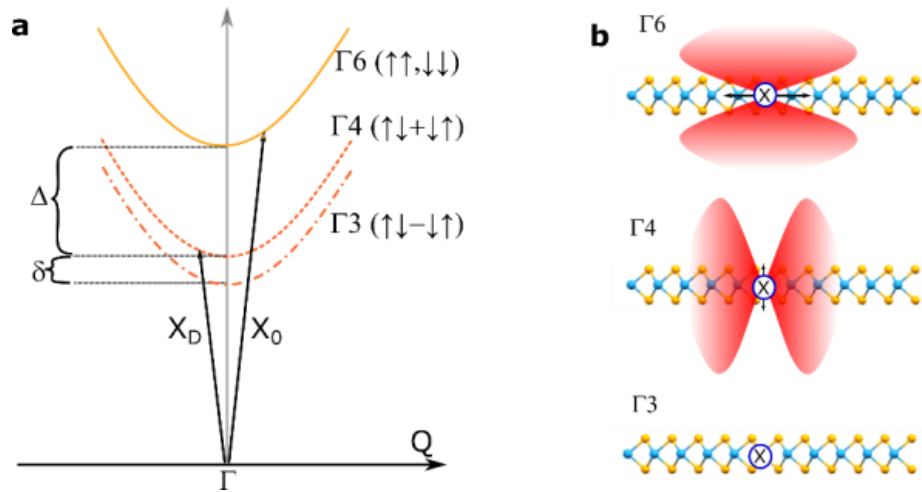

**Figure 1 | Schematic drawing of the excitonic non-degeneracy and the dark exciton landscape in tungsten diselenide. a** Possible bright and dark exciton states for neutral WSe$_2$ arising from the crystal symmetry, indicated in the two-particle picture around zero momentum ($\Gamma$ point). The bright exciton ($X_0 - \Gamma 6$) is separated from the grey exciton ($X_{D,g} - \Gamma 4$) and the dark exciton ($X_D - \Gamma 3$). Arrows indicate the optical excitation of the respective states. In contrast to bright excitons, the grey exciton is only dipole allowed for z-polarization (corresponding to an out-of-plane dipole). A sketch of the expected radiation pattern for the different symmetries respectively excitons is shown in **b**.

**Results**

The angle-resolved photoluminescence measurement has been carried out on an h-BN encapsulated WSe$_2$ ML. An atomic-force microscopy (AFM) image of the assembled stack is shown in **Figure 2a**. Experiments were performed under pulsed quasi-resonant excitation with an effective detuning of 54 meV to the bright exciton. A schematic of the detection concept is presented in **Figure 2b**.
The corresponding two-dimensionally plotted (2D) PL spectrum (energy as a function of emission angle) can be seen in **Figure 3a**, which is displayed in false-colour linear intensity scale (white: minimum, dark blue: maximum). Several PL peaks can be identified, which are attributed to a variety of excitonic species, ranging from neutral exciton ($X^0$), charged species ($X^-$), biexciton (XX) to grey ($X^0_{D,g}$) and dark ($X_D$) excitons. Some peaks feature a fine structure, which is taken into account, whereas some peaks arise from acoustic (ac) or optical (op) phonon sidebands (SB), as labeled in the line spectrum in **Figure 3b**. For clarity and linewidth analysis, the angle-integrated spectrum has been fitted with a sum of Lorentzian curves (see **Figure 3b**), which can be used to describe nearly-homogeneously broadened excitonic lines. For the sake of comparison, all species have been fitted with the same line profile. The obtained line parameters are summarized in Table 1. The peak positions above 1.67 eV are in agreement with Barbone et al.[5] ($X^0$, $XX^0_1$, $XX^0_2$, $X^-_{inter}$, $X^-_{intra}$, $X_{D,g}$, $XX^-$) and Chen et al.[6] ($X^0$, $XX^0$, $X^-_{inter}$, $X^-_{intra}$, $X_{D,g}$, $XX^-$), while the low energy feature can be explained by the predicted phonon-assisted sideband[13] emission from the dark excitons (K-K' and K-$\Lambda$ transitions). As encapsulation is known to change the band structures as well as the exciton binding energies, above comparison are only done with similarly encapsulated samples. The bright–grey splitting is extracted from the spectrum with $\Delta = 43$ meV. Taking the phonon-band structure from Terrones et al.[22] (with phonon wavenumbers LO 260 cm$^{-1}$, TO 2 cm$^{-1}$, LA 125cm$^{-1}$ TA 100 cm$^{-1}$) into consideration, the energetic position of the dark exciton arising from the K-K' transition can be calculated as 1.681 eV and the one from the K-L transition as 1.690 eV, which are in good agreement with predicted values from Brem et al.[13]. To further confirm the identification as phonon sidebands, a temperature series has been performed (see Fig. SI.4), the experimental data of which well resembles the prediction of Brem et al.[13].

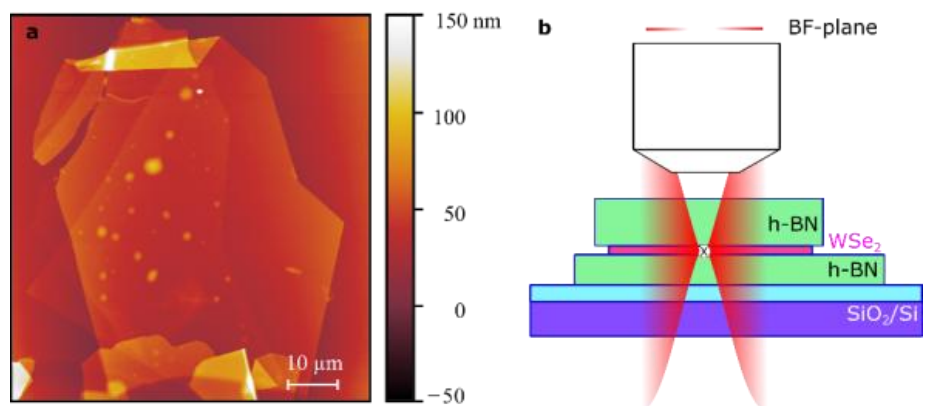

*Figure 2 | Sample image and measurement concept. a AFM measurement of the sample and b conceptual drawing of the measurement showing a schematic radiation pattern for an in-plane radiation from the exciton and the resulting expected intensity distribution in the back-focal (BF) plane of the microscope objective.*

The grey exciton can be easily identified in the angle-dependent spectrum, as it shows a clearly deviating radiation pattern (for a schematic see **Figure 2b**). Its position agrees with aforementioned studies[5,6,13,17]. In such 2D spectrum, bright excitons feature an almost constant intensity for all collected angles, while the rather flat profile shows slightly reduced intensities at higher angles. In contrast, the grey exciton is hardly present at 0° emission angle (normal to the sample surface) but exhibits a drastic increase in intensity towards the detection limit of emission angles due to the finite numerical aperture of the objective used in the experiment. A minor tilt in the sample plane with respect to the objective normal even allows us to detect an angle span of approximately +41 to -31°.

To highlight the clear differences in the radiation pattern, the corresponding PL intensity as a function of the emission angle is shown as a polar plot for both the grey and bright neutral exciton in **Figure 3c**. The intensity levels are in relation to **Figure 3a** and due to a varying background signal level in the range of different densely-packed species, the two species can show different intensity levels at 0°. Astonishingly, one can unambiguously identify the grey exciton due to its expected behavior of radiating in the WSe₂ plane instead of perpendicular to the monolayer. This does not only provide experimental evidence of such a Γ4 species or give a tool at hand to distinguish them from different excitonic modes, but it also verifies the prediction made by group-theory analysis (cf. Robert et al.[15]). In fact, all other excitonic modes such as trions, biexcitons and phonon sidebands show a similar pattern as the representative neutral bright exciton.

*Table 1|* **Extracted parameters of the PL peaks obtained under quasi-resonant pulsed excitation.** A multi-peak fit with Lorentzian line profiles yields the following central energies of different excitonic (X) species and their full width at half maxima (FWHM). The corresponding spectrum with labeled features is shown in **Figure 3b**.

| Peak | $X^0$ | $XX^0_1$ | $XX^0_2$ | $X^-_{inter}$ | $X^0$ op.SB | $X^-_{intra}$ | $X_{D,g}$ | $XX^-$ | $X_D$ (K-K' TA) | $X_D$ (K-K' LA) | $X_D$ (K-Λ TO) | $X_D$ (K-Λ LO) | $X_D$ (K-K' TO) | $X_D$ (K-K' LO) |
|---|---|---|---|---|---|---|---|---|---|---|---|---|---|---|
| Energy (eV) | 1.729 | 1.712 | 1.709 | 1.699 | 1.696 | 1.691 | 1.686 | 1.678 | 1.670 | 1.666 | 1.663 | 1.655 | 1.651 | 1.646 |
| FWHM (meV) | 5.0 | 2.2 | 4.0 | 4.0 | 3.7 | 3.9 | 3.0 | 4.0 | 11.0 | 5.0 | 2.4 | 10.0 | 8.0 | 8.0 |

From first sight it is clear, that the measured patterns do not resemble the well-known dipole radiation pattern. The modification can be a consequence of the surrounding dielectrics and occurring interferences in a multilayer structure. However, in order to verify that the distinct patterns really arise from an in- and out-of-plane dipole, an electromagnetic simulation was done to calculate the farfield pattern. Hereby, the anisotropy of h-BN and WSe$_2$ was explicitly taken into account (for further details we refer to the Methods section). Indeed, as can be seen by **Figure 3c**, the simulated pattern and the measured pattern are in good agreement with each other. Generally, it can be stated that, for both resonances, the main lobe is surprisingly not mostly directed to the substrate. While being modified by the encapsulation, the difference between the two radiation patterns is still striking. The detailed analysis of the polarization from the simulation can be found in the Supporting Information together with a co- and contra-polarized PL measurement. While the general radiation patterns are not changing, the grey exciton can experimentally only be seen in a cross-polarized measurement, similar to the simulation (cf. **Figs. SI.1, SI.2 and SI.3**).

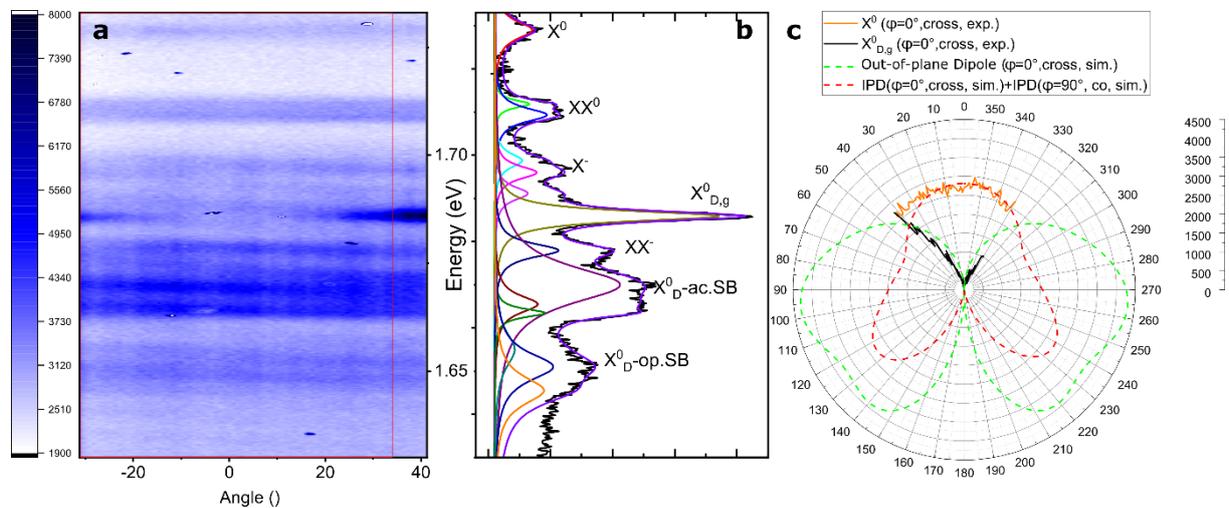

*Figure 3 | Angle-resolved PL spectrum and emission pattern. a* *2D false-colour contour plot of an angle-resolved PL spectrum of the encapsulated WSe$_2$ monolayer under pulsed excitation at 1.789 eV with an excitation density of 78 µJ/cm$^2$ at 10 K.* ***b*** *Corresponding angle-integrated line spectrum (from red-boxed region in **a**) showing several excitonic features as well as excitonic phonon sidebands. A fit with multiple Lorentzian curves is shown on top of the measured spectrum (dotted curve) with underlying individual peaks labeled (solid curves).* ***c*** *Comparison of the experimentally obtained radiation patterns for both the bright and the grey exciton to the simulated results. For the in-plane dipole (IPD), the contribution of dipoles perpendicular to the excitation arising from valley dephasing were considered in the simulation, too. Therefore, a degree of linear polarization of 5% was determined from measurements (Fig. SI.3). All intensity scales are linear.*

**Conclusion**

The photoluminescence of h-BN-encapsulated WSe$_2$ was analyzed by means of angle-resolved PL spectroscopy. A rich spectrum with numerous excitonic features was obtained that agrees well with previous predictions and measurements on high-quality samples. Strikingly, angle-resolved measurements allow one to clearly distinguish in-plane emitting from out-of-plane emitting excitons, as the analysis of the radiation patterns from excitons shows in agreement with electrodynamic simulation. While most of the features show almost no angle dependence, the grey exciton's signal rises drastically towards higher angles, as predicted for this species. This provides a unique tool for both monolayer samples as well as multilayer stacks, in which various intra- and interlayer excitonic features can form particularly at cryogenic temperatures. This motivates further studies involving charge transfer excitons or hybridized states with partial charge transfer, where a change of dipole moment direction is expected as well.

**Methods**

**Sample fabrication.** Tungsten diselenide (WSe$_2$) bulk single crystals were grown in an excess selenium flux (defect density: 5 x 10$^{10}$/cm$^2$). For encapsulated samples, monolayer WSe$_2$ and h-BN were first exfoliated from bulk single crystals onto SiO$_2$. For WSe$_2$, the SiO$_2$ substrate was first exposed to an O$_2$ plasma step before exfoliation. Monolayers and thin h-BN were both identified by optical contrast using a microscope. Afterwards, a dry stacking technique using polypropylene carbonate (PPC) on PDMS was used to pickup and stack h-BN/WSe$_2$ layers. First a top layer of h-BN (~20 nm) is picked up at 48 degrees C, then WSe$_2$, and finally the bottom layer of h-BN (~30 nm). After each h-BN pickup step the PPC is briefly heated to 90 C to re-smooth the PPC and ensure a clean wave front. For transferring the stack onto a clean substrate, the substrate is first heated to 75 degrees C, the stack is then put into contact, and gradually heated to 120 degrees C. Afterwards, the PPC/PDMS is lifted and the substrate is immersed in chloroform and rinsed with IPA to remove polymer residue. Atomic-force microscopy confirmed a total stack thickness of ~40 nm (~10 nm + ~20 nm for the encapsulants).

**PL measurement.** The measurements have been done using a conventional 4f µ-PL setup with confocal selection. The sample was mounted in a continuous-flow cryostat at high vacuum and was cooled down to 10K. A 40x (NA 0.6) microscope objective has been used to focus a pulsed Titan-Sapphire laser at 1.789 eV onto the sample. A short-pass filter for 700 nm was used to shape the pulse in front of the sample. A long-pass filter 700 nm and a polarizer after the sample were used to suppress the laser in the collection path of the PL signal, which was detected by a monochromator with nitrogen-cooled camera.

**Simulations.** The farfield pattern of the given structure was simulated using CST microwave studio. The thicknesses for the simulation were taken from the AFM measurement of the structure. Furthermore, the anisotropic refractive index of h-BN has been taken from Segura et al. [23] and for WSe$_2$ a hybrid approach was taken. The out of plane refractive index was taken from Laturia et al. [24], $\epsilon_{\infty,z} = \epsilon_{s,z} = 7.5$. For the in-plane permittivity, a Lorentz model was employed to account for the resonance of the bright A-1s transition. Here, the following values were used: $\epsilon_{\infty,x,y} = 15, \epsilon_{s,xy} = 15.22$ and a damping frequency of 4.77 THz. The contribution of the excitons, especially the grey exciton, to the permittivity in out of plane direction is about 1000 weaker than for the in-plane component[17,25], giving no significant contribution to the permittivity. Therefore, this contribution was neglected for the simulation. For silicon oxide and silicon, they were taken from the programs database. The resulting farfield patterns were analyzed in terms of polarization by projecting them on the unit vector of the radiation sphere using the Ludwig 3 convention.

**Visualization.** The schematic depiction of the WSe$_2$ monolayer in Fig. 1b is based on crystallographic data provided by the *Materials Project*[26] and drawn by the tool *Mercury*[27].

**Acknowledgement**

The authors acknowledge financial support by the German Research Foundation (DFG: SFB1083, RA2841/5-1), by the Philipps-Universität Marburg, and the German Academic Exchange Service (DAAD). Synthesis of WSe$_2$ and heterostructure assembly are supported by the NSF MRSEC program through Columbia in the Center for Precision Assembly of Superstratic and Superatomic Solids (DMR-1420634). The authors thank G. Witte and D. Günder for assistance with the AFM measurement.

**Authors' contributions**

A.R.-I. conceived the experiment and initiated the study on angle-dependent measurements in 2015. The joint work was guided by J.C.H. and A.R.-I. High-quality WSe$_2$ synthesis and heterostructure assembly were achieved by S.S.E., D.A.R., K.B. and J.C.H. The setup was established by L.M.S. and A.R.-I., and the structures measured by L.M.S. The results were interpreted, discussed and summarized in a manuscript by L.M.S. and A.R.-I. with the support of all coauthors.

**Corresponding author**

Lorenz Maximilian Schneider: maximilian.schneider@physik.uni-marburg.de

Arash Rahimi-Iman: a.r-i@physik.uni-marburg.de

**Authors' statement/Competing interests**

The authors declare no conflict of interest

# Supporting Information

## Direct Measurement of the Radiative Pattern of Bright and Dark Excitons and Exciton Complexes in Encapsulated Tungsten Diselenide

Lorenz Maximilian Schneider[1,*], Shanece S. Esdaille[2], Daniel A. Rhodes[2], Katayun Barmak[3], James C. Hone[2], and Arash Rahimi-Iman[1,*]

[1]Faculty of Physics and Materials Sciences Center, Philipps-Universität Marburg, Marburg, 35032, Germany

[2]Department of Mechanical Engineering, Columbia University, New York, NY 10027, USA

[3]Department of Applied Physics and Applied Mathematics, Columbia University, New York, NY 10027, USA

## S1. Polarization analysis

The simulation was once performed for an in-plane and out-of-plane dipole located in the middle of the monolayer, respectively. The far-field patterns were calculated at different spectral positions, i.e. that of the bright and the grey exciton, as displayed in. **Figure SI.1**. One can see that for this structure, the patterns are indeed modified; nevertheless, the emission from in- and out-of-plane dipoles can be still easily distinguished. While the in-plane dipole has a well-defined orientation in the simulation, in experiment data the emission is expected to be found for a preferred direction of microscopic dipoles (i.e. along the lasers polarization), but not exclusively for that direction owing to valley dephasing. This is especially important for the here used rather strong pulsed excitation, which favours these dephasing processes.

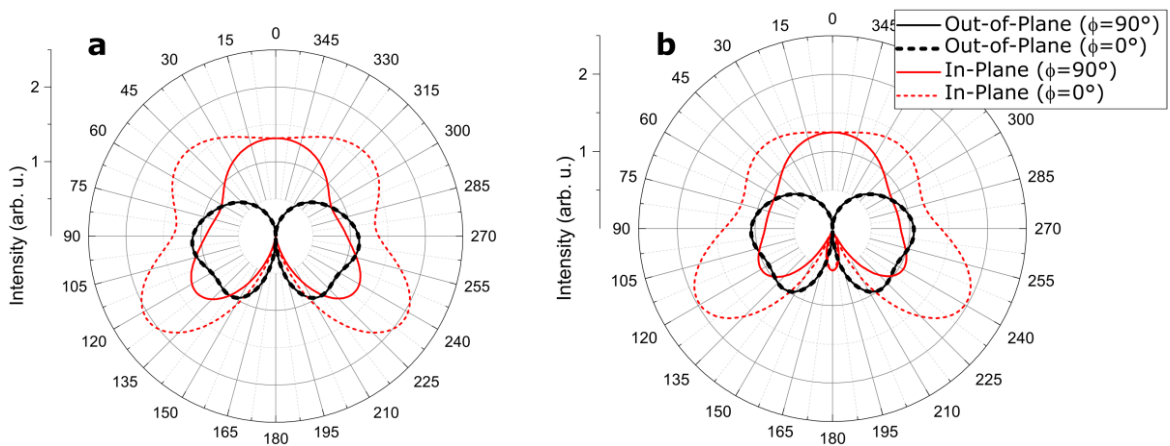

**Figure SI.1 | Simulated radiation patterns**. Farfield intensity patterns simulated for an electronic dipole emission at the spectral position of the bright exciton (a) as well as the grey/dark exciton (b) for both linear polarizations. Both polar diagrams show the radiation patterns for an in-plane (located along the x-axis) and an out-of-plane dipole, in the x-z (ϕ=0°, dashed line) and y-z (90°, solid line) planes. The pattern for the out-of-plane dipole has been multiplied by 100 to show all data on the same intensity scale, whereas the profiles for 0 and 90° are congruent. The 90-270° line corresponds to the monolayer orientation, with angles between those values representing emission to the substrate side.

This means that a weighted average over both cases, i.e. parallel and perpendicular to the dipole, is expected in experiment. For the out-of-plane dipole, however, we find no dependence of the emission pattern to the selected plane (x-z or y-z).

In a next step, the polarization was evaluated by projecting the polarization of the beams emitted from a dipole to the unit-vectors of a sphere according to the Ludwig 3 convention for the relevant cases (see **Figure SI.2**). For the in-plane dipole, the simulation shows that the pattern indeed varies with the polarization, i.e. detection co- (black) or cross-polarization (red). Nevertheless, as the difference in magnitude between co- and cross polarization is strong (from 2.5 to 8 orders of magnitude), the contribution of dipoles oriented 90° with respect to the laser polarization arising from valley decoherence will be more significant. For the out-of-plane dipole, it is remarkable that for a fixed detection plane, e.g. for φ=0°, the radiation is only found for cross-polarized emission. Orientations with negligible value level (<<1E-5) are neglected in the polar plots.

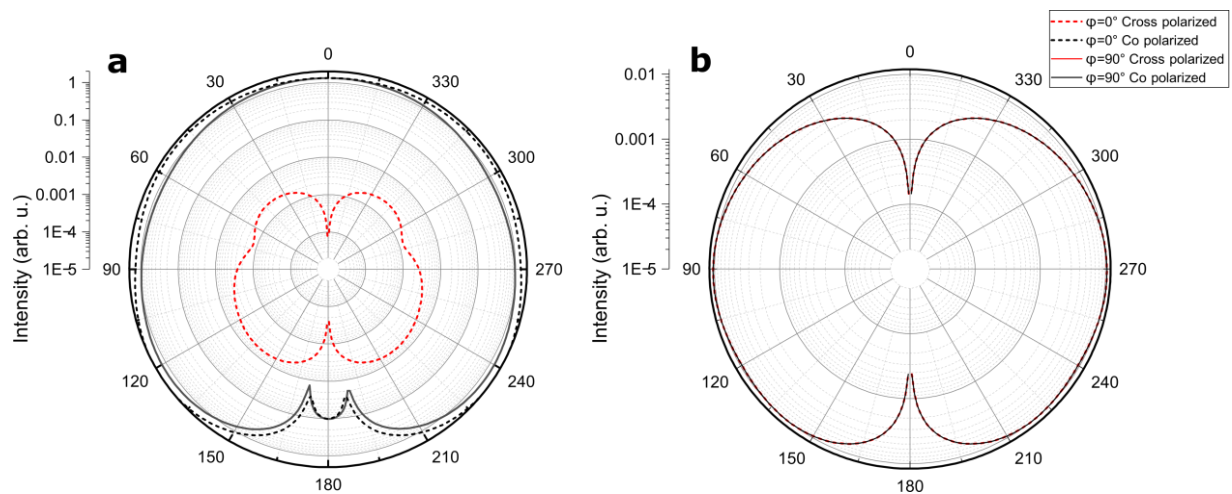

**Figure SI.2 | Analysis of the polarization for the simulation results**. Intensity profile of detected signal for co-/cross-polarized light in the plane parallel (φ=0°), as well as perpendicular (φ=90°), to the dipole. The analysis has been performed for an in-plane dipole at the spectral position of the bright exciton (a), as well as for an out-of-plane dipole at that of the grey exciton (b).

Furthermore, to check the polarization dependence of the emission pattern experimentally, an angle-resolved measurement has been performed under linearly co- and cross-polarized detection (see **Figure SI.3**).

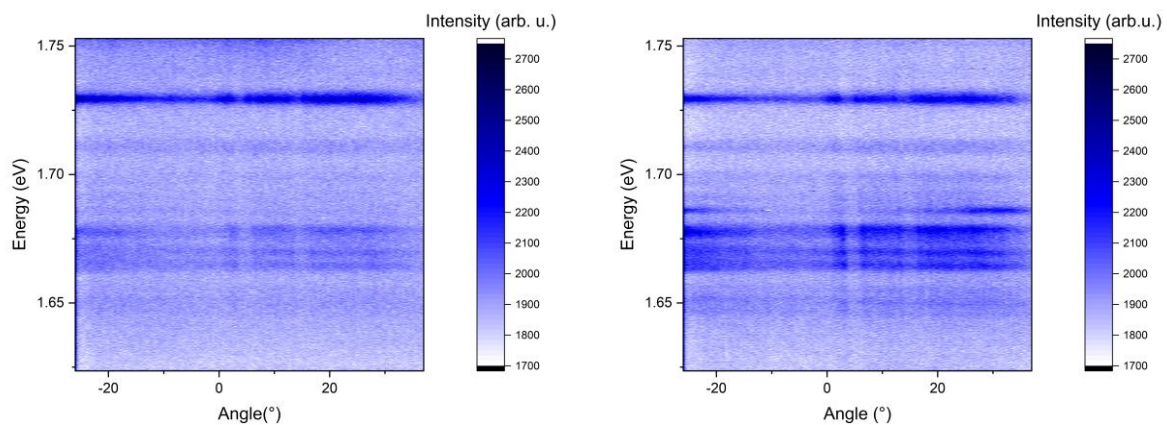

**Figure SI.3 | Co- and cross-polarized emission spectra.** Angle-resolved measurements of photoluminescence under co- and cross-polarization (left and right, respectively) in the plane parallel to the laser excitation (φ=0°).

The most remarkable difference between these measurements is the absence of the grey exciton under co-polarized emission, which is in good agreement with simulated predictions for the selected detection plane.

## S2. Temperature Series

A temperature series performed on a high-quality hBN/WSe2/hBN stack on a SiO2/Si substrate verifies the presence of phonon side bands (see **Figure SI.4**). A comparison to theoretical predictions of Ref. [13] reveals that the experimental temperature dependence of the identified phonon sidebands is in good agreement with calculations, as the contour diagram with normalized spectra and relative energies in **Figure SI.4** shows.

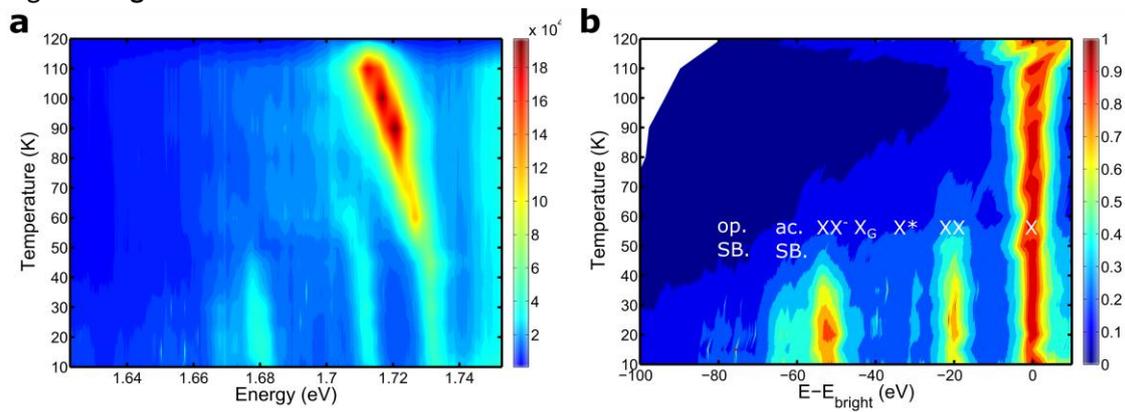

**Figure SI.4 | Temperature-dependent photoluminescence.** Raw µ-PL data of the temperature series presented in contour diagram stitched from individual line spectra (a). Temperature-dependent spectra normalized and plotted relative to the exciton energy (b). Here, both the representation as well as the signatures' behavior is similar to that in the literature concerning theoretical predictions[13] (not shown here).